\documentclass[aps,prl,epsfigure,twocolumn,superscriptaddress]{revtex4-1}
\usepackage{dcolumn}    
\usepackage{bm} 
\usepackage{graphicx}
\usepackage{amsmath}    
\usepackage{latexsym}
\usepackage{amsfonts}   
\usepackage{amssymb}
\usepackage{array}      
\usepackage{epsfig}
\usepackage{txfonts}
\usepackage{color}
\usepackage[colorlinks=true,linkcolor=blue,urlcolor=blue,citecolor=blue]{hyperref}
\usepackage{enumerate}
\usepackage{bbold}
\usepackage{epstopdf}

\def \ell{{d}}

\newcommand{\beginsupplement}{%
	\setcounter{table}{0}
	\renewcommand{\thetable}{S\arabic{table}}%
	\setcounter{figure}{0}
	\renewcommand{\thefigure}{S\arabic{figure}}%
}

\newcommand{\cnr}{CNR--SPIN, Dipartimento di Scienze Fisiche, Universit\`a di 
Napoli Federico II, I-80126, Napoli, Italy}

\newcommand{\majulab}{MajuLab, CNRS-UCA-SU-NUS-NTU International Joint Research 
Unit, Singapore}
\newcommand{\ntu}{Division of Physics and Applied Physics, School of Physical 
and Mathematical Sciences, Nanyang Technological University, Singapore}

\begin{document}
\author{Yuanjian Zheng} 	
\affiliation{\ntu}
\author{Yan-Wei Li}
\affiliation{\ntu}
\author{Massimo Pica Ciamarra} 
\affiliation{\ntu}
\affiliation{\majulab}
\affiliation{\cnr}
\title{Hyperuniformity and generalized fluctuations at Jamming}

\begin{abstract}        
The suppression of density fluctuations at different length scales is 
the hallmark of hyperuniformity. However, its existence and significance in jammed solids is still a matter of debate. We explore the presence of this hidden order in a manybody interacting model known to exhibit a rigidity transition, and find that in contrary to exisiting speculations, density fluctuations in the rigid phase are only suppressed up to a finite lengthscale. This length scale grows and diverges at the critical point of the rigidity transition, such that the system is hyperuniform in the fluid phase. This suggests that hyperuniformity is a feature generically absent in jammed solids. Surprisingly, corresponding fluctuations in geometrical properties of the model are found to be strongly suppressed over an even greater but still finite lengthscale, indicating that the system self organizes in preference to suppress geometrical fluctuations at the expense of incurring density fluctuations. 
\end{abstract}
\maketitle
The study of spatial structure is an important aspect of understanding the behavior of disordered matter. However, unlike crystals, disordered systems possess no obvious long range order or lattice symmetries that severely limit the available tools at hand to characterize them. For this reason, the concept of hyperuniformity which specifies the anomalous suppression of density fluctuations becomes an essential construct in understanding disordered media \cite{Torquato2018}. For a system consisting of $n$ particles located at positions $\vec{r}_i$ in a sample material, the number of particles that fall within an observation window of radius $R$ centred around a point $\vec{r}_c$ can be written
\begin{equation}
 N(R)=\sum^n_{i} \Theta ( R - \vert \vert \vec{r}_i-r_c \vert \vert )
\end{equation}
where $\Theta(x)$ is the heaviside function. For a given $R$, the variance in $N$ as $\vec{r}_c$ is shifted around the sample characterizes the density fluctuations and is simply given by: 
\begin{equation}
	\langle \delta^2 N  \rangle = \langle N(R)^2 \rangle- \langle N(R) 
\rangle^2
	\label{eq:number_fluctuation}
\end{equation}
For a random distribution of points, the density fluctuations scale as the volume of the observation window,$\langle \delta^2N \rangle \sim R^d$, while a scaling with surface area:  $\langle \delta^2N \rangle \sim R^{d-1} $, is a feature of ordered crystals and quasi-crystals, where $d$ here denotes the spatial dimension of the system. Disordered hyperuniformity is thus a description of spatially disordered systems where density fluctuations grow with $R$ at a rate slower than that of a uniformly random distribution of points, i.e. $\langle \delta^2N \rangle \sim R^{\epsilon} $ where $ 1 \leq \epsilon \leq 2 $ in the limit of $R \to \infty$, with the extreme suppression of fluctuations reminiscent to behavior of crystals termed perfect hyperuniformity ($\epsilon=1$) \cite{Torquato2018,TorquatoStilinger2003}. As such, disordered hyperuniformity is often regarded as a form of hidden order in nature. 

While difficult to visually identify \cite{Torquato2018}, hyperuniformity in density fluctuations of point distributions is a seemingly recurring theme of nature. Examples include the distribution of photoreceptors in avian retina \cite{JiaoTorquato2014}, self-organization of immune systems \cite{MayerWalczak2015}, ordering of prime numbers \cite{TorquatoCourcyIreland2018} and in ground state properties of complex systems \cite{TorquatoStillinger2015}. Furthermore, hyperuniform materials are also being synthesized for their potential applications, such as with the nanofabrication of photonic materials \cite{ManSteinhardt2013} or its use in surface-enhanced Raman spectroscopy \cite{DeRosaSasso2015}. However, despite its prevalence, the exact role of hyperuniformity in many such systems remain unclear, understanding of its origins limited \cite{HexnerLevine2015,HexnerLevine2017}, and in certain instances, even its existence is still debated upon. 

One such debate occurs over the existence of hyperuniformity in jammed solids, disordered assembly of particles interacting via short-ranged repulsive forces 
that are mechanically rigid due to a sufficiently high packing density \cite{LiuNagel2010}. In what is known as the Jamming-Hyperuniformity conjecture, strictly jammed systems defined by the complete absence of rattlers, particles that are locally unconnected to the rest of the material, are postulated to be always hyperuniform in finite dimensions  \cite{Torquato2018}. While there is some indication that this might be true \cite{ConiglioAste2017}, it is  not without considerable protest \cite{IkedaBerthier2015,IkedaParisi2017, GodfreyMoore2018}, to which Torquato in \cite{Torquato2018} attributed to an inadequate consideration for the rattlers,  which violates the condition of ``strictly" in the conjecture. Regardless, even if hyperuniformity is indeed present in all strictly jammed solids, the question of what exactly is its role in jamming remains. Is this connection causal or coincidental? And in the broader picture, what is the connection of hyperuniformity with the underlying systems in which it emerges from, especially given that they are all of such diverse nature?

\begin{figure*}
		\includegraphics[width=\textwidth]{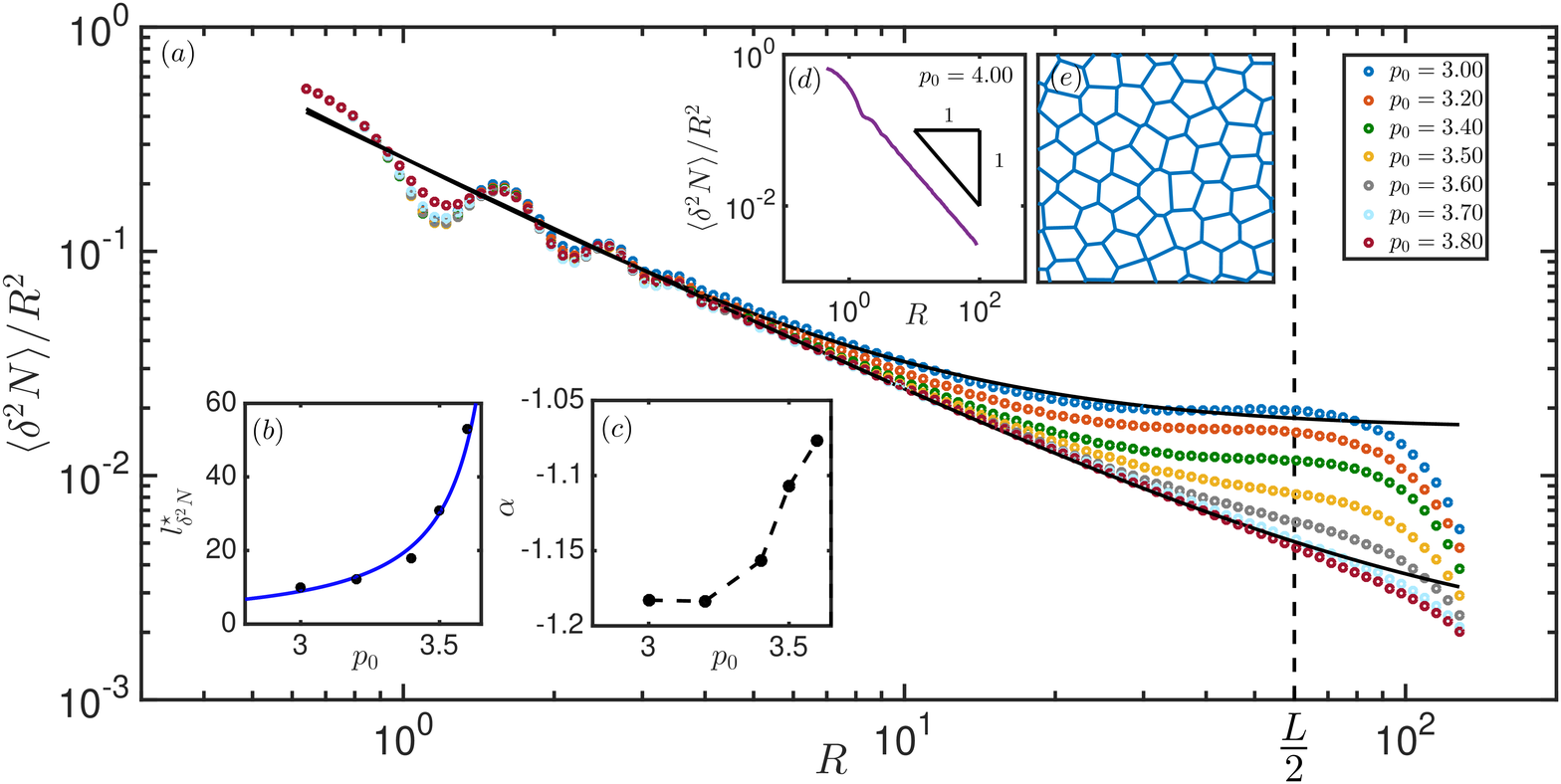}
		\caption{(a) $\langle \delta^2 N \rangle /R^2$ as a function of $R$ at various values of $p_0$ in the rigid phase ($p_0 < p_0^{\star}$) for a system size of $N=16384$ particles. $\langle \delta^2 N \rangle /R^2$ exhibits a continuous crossover from power law behavior at shorter length scales where density fluctuations are strongly supressed to a behavior that corresponds more to random configurations at longer length scales. The black dashed line indicates a cut-off length scale of $L/2$ where finite-size effects are prominent for larger values of  $R$ (See Supplementary Material).The black solid lines represent the best fit of the scaled number fluctuations to the ansatz - $\langle \delta^2 N\rangle /R^2 \sim kx^{\alpha}+\beta$, for $R < L/2$, and where $l^{\star}_{\delta^2 N} \equiv (\beta/k)^{1/\alpha}$  (b,c) $l^{\star}_{\delta^2 N}$ and $\alpha$ extracted from the fit in (a).  (b) the blue line represents a fit of $l^{\star}_{\delta^2 N} \sim \vert \vert p_0-p_0^{\star}\vert \vert^{-\nu} $ with $p_0^{\star}=3.813$ and $\nu$ found to be $1.275$, consistent with the picture of a diverging lengthscale at the critical point of the rigidity transition. (d) Density fluctuations in the fluid $(p_0=4.00)$ are strongly suppressed and the system is perfectly hyperuniform (e) A partial snapshot of a Voronoi tessellation at $p_0=3.20$.
} \label{fig:number_fluctuations}
	\end{figure*}

To begin addressing these questions, we consider a biologically motivated model of cells in an epithelial monolayer that is well known to exhibit a rigidity 
transition \cite{BiManning2015,BiManning2016}. In this model, particles are distributed on a 2D surface and a cell is attributed to each particle $i$ by a 
Voronoi tessellation of the space, a confluent partitioning of the plane that by construction do not possess any rattlers, which we highlight in  Fig. 
\ref{fig:number_fluctuations}(e). From each prescribed cell, geometrical properties of area $A_i$ and perimeter $P_i$ is attributed to the corresponding 
particles that in turn completely defines a non-dimensional energy of the monolayer that is derived from biophysical considerations of the underlying single cell mechanics  \cite{FarhadifarJulicher2007,StapleJulicher2010,BiManning2015} 
\begin{equation}
E=\sum_i E_i \sim \sum_i (A_i-1)^2+\frac{(P_i-p_0)^2}{r}.
\end{equation}
The control parameter $p_0 \equiv P_0/ \sqrt{A_0}$ of this local energy functional is regarded as a target shape index given by the ratio of a given preferred perimeter $P_0$ to an effective length scale associated to the corresponding preferred area $A_0$, and thus effectively controls local stiffness of each particle \cite{LiPicaCiamarra2018, MosheMarchetti2017}, while $r$ defines  the relative contributions to the energy functional attributed to deviations from respective preferred values of the two geometrical components, where we consider $r=1$ hereafter for simplicity. In the thermodynamic limit, this parameter $p_0$, governs melting \cite{LiPicaCiamarra2018} and rigidity transitions \cite{BiManning2015,BiManning2016}. In particular, at zero temperature and in the absence of activity, the system undergoes a solid to fluid transition at the critical point $p_0^{\star}=3.813$ in the direction of increasing $p_0$ \cite{BiManning2015,BiManning2016}.

Here we note that in this model, interaction is truly manybody, in that the mechanical forces ($\vec{F}_i = - \nabla \vec{E}_i$ ) acting on the degrees of freedom are not reducible to pairwise forces by virtue of the manybody energy functional \cite{BiManning2016}, a key feature that is not shared by conventional models used to study the jamming transition \cite{LiuNagel2010}, especially in context of hyperuniformity \cite{Torquato2018}. In the following, we first examine the density fluctuations of energy minimal (metastable) configurations as a function of the parameter $p_0$, and show that it is in fact not hyperuniform in the long wavelength limit. In particular, density fluctuations are suppressed only at finite length scales in the rigid phase, and seem to diverge  at the critical point of the rigidity transition, such that the system is always hyperuniform in the fluid phase $(p_0 > p_0^{\star})$. 
Subsequently, we then show how a generalized form of fluctuations in this model is suppressed over an extended length scale that is greater than that of the underlying density fluctuations, indicating the presence of self organization; a preference for, or drive towards the suppression of fluctuations beyond its density. 

To obtain the energy minimal metastable configurations of the model for hyperuniform analyses, we utilize a conjugate gradient algorithm \cite{GalassiGNU} starting from random initial conditions of the point distribution. Unless otherwise stated, each energy minimal state is given by 2D Voronoi tessellations of $N=16384$ particles in a square box of length $L=\sqrt{N}$ under periodic boundary conditions, such that the system is always confluent with mean area $\bar{A}=1$. Fluctuation statistics are obtained from averaging $25$ unique realizations at any given value of $p_0$, and $10^4$ observation windows for each value of $R$ of each realization. In addition, we have also checked that the qualitative picture does not change if the energy minimization protocol is performed using overdamped dynamics 
\cite{BiManning2016}. 

In Fig.\ref{fig:number_fluctuations}(a) and (d) we show the scaled density fluctuations $\langle \delta^2 N\rangle / R^2$ as a function of $R$ for various $p_0$ in the rigid ($p_0 < p_0^{\star}$) and fluid regimes ($p_0=4.00$) respectively. In the rigid regime, the system exhibits anomalous suppression of density fluctuations consistent with hyperuniformity ($\epsilon < 2$) at shorter length scales. However, we see that at some value of $R$ dependent on the value of $p_0$, the system exhibits a continuous crossover to a plateau in its scaled fluctuations ($\epsilon=2$), suggesting that it is in fact not hyperuniform. At even larger $R$ the fluctuations decay due to finite size effects, which we show in the Supplementary Material. 
Here, we note that the study of hyperuniformity in this model has very recently been examined in context of generating photonic band gaps as an approach to designing novel materials \cite{LiBi2018}. However, the smaller system sizes considered were insufficient (See Supplementary Material - Fig. \ref{fig:finite_size}) to probe the longer wavelength behavior of the system and thus Li et al. were unable to conclude on the true extent of fluctuation suppression in the density. Indeed, here we find that while density fluctuations are anomalously suppressed at shorter length scales of the system, there exists a crossover of the density fluctuation to a scaling more reminiscent of random configurations at larger length scales, representing a significant departure from the key assumption in \cite{LiBi2018}.

Now, by assuming the following ansatz for the crossover in fluctuations: $\langle \delta^2 N \rangle /R^2 \sim kx^{\alpha} + \beta$, we perform a fit for the region smaller than a cut-off lengthscale indicated by the black-dashed line (Fig. \ref{fig:number_fluctuations}(d)) corresponding to half of the system size, such as to mitigate the influence of finite size effects and extract a $p_0$ dependent lengthscale $l^{\star}_{\delta^2 N}$ defined:
\begin{equation}
l^{\star}_{\delta^2 N}(p_0) \equiv  (\beta/k)^{1/\alpha}
\end{equation}
We find that $l^{\star}_{\delta^2 N}$ is indeed monotonically increasing with $p_0$ (Fig. \ref{fig:number_fluctuations}(b)), and a subsequent fit of $l_{\delta^2 N}^{\star}$ (blue solid line) to the scaling ansatz: $l^{\star} \sim \vert\vert p_0-p_0^{\star} \vert\vert^{-\nu}$ with $p_0^{\star}=3.813$ held fixed, yields $\nu=1.275$. This indicates that the length scale extracted from $\langle \delta^2 N \rangle$ is consistent with its divergence at the known rigidity transition \cite{BiManning2015,BiManning2016}. The power law index $\alpha$, corresponding to the suppression of $\langle \delta^2 N \rangle$ within the length scale $l^{\star}_{\delta^2 N}$, is also plotted in Fig. 
\ref{fig:number_fluctuations}(c). We note that $l^{\star}_{\delta^2 N}$ extracted for larger values of $p_0$ are greater than the cut-off lengthscale and are also in a region where energy minimization is known to be difficult \cite{BiManning2015}. This critical slowing down of the dynamics has 
been known to influence the correct interpretation of analyses on hyperuniformity \cite{AtkinsonTorquato2016}, but we stress that our results remain robust and consistent with the observed rigidity transition. 

In fact, we show that perfect hyperuniformity in the density fluctuations is recovered for configurations deep in the fluid phase ($p_0=4.00$) even where energy minimization is never truly completed, since $l_{\delta^2 N}^{\star}$ is expected to have already diverged. Here we  also note that, hyperuniformity is indeed expected in the fluid phase since both the probability density of the area and of the perimeter of the cells approach a delta function. This in turn represents a qualitative departure from conventional discussions of  hyperuniformity at the Jamming transition, since hyperuniformity even if present either in the jammed phase and / or at the critical point, is not traditionally a behavior associated to the fluid phase of the system. In this regard, it is worth emphasizing that the fluid phase of the Voronoi model is qualitatively distinct from the unjammed phase of sphere packings; While both models have zero shear moduli, unjammed packings of spheres are locally unjammed\cite{TorquatoStillinger2010}, while the fluid phase of the Voronoi model is locally jammed, as all particles are in a local minimum of the energy. One might thus speculate that hyperuniform packings are locally jammed but globally unjammed systems.

We now consider fluctuations of other quantities of the model by generalizing the notion of hyperuniformity to scalar fields \cite{Torquato2016,MaTorquato2017}.
We note a recent approach that involves generalized fluctuations that identified diverging lengthscales associated to jamming based on fluctuations of the local coordination number $Z$, the average number of contacts per particle \cite{HexnerNagel2017}. The fluctuations in $Z$ are found to be more suppressed than that of the underlying density, which indicates that Jammed systems do suppress fluctuations of different quantities to different extent. But why is this so? And what is the quantity in which fluctuations are most suppressed?

The fluctuations in $Z$ are of little physical relevance to the model at hand, since Voronoi topologies are often over coordinated and contacts do not reflect the physical interactions that are inherently manybody and are irreducible to pairwise interactions in nature. Instead we study the fluctuations associated to the geometrical properties of the prescribed cell, that enter the energy functional explictly
\begin{equation}\label{eq:energy}
X(R)=\sum^n_{i} X_i \Theta ( R - \vert \vert \vec{r}_i-r_c \vert \vert )
\end{equation}
where $X= $ $A$ or $P$, for the area and perimeter respectively.
\begin{figure*}
\includegraphics[width=\textwidth]{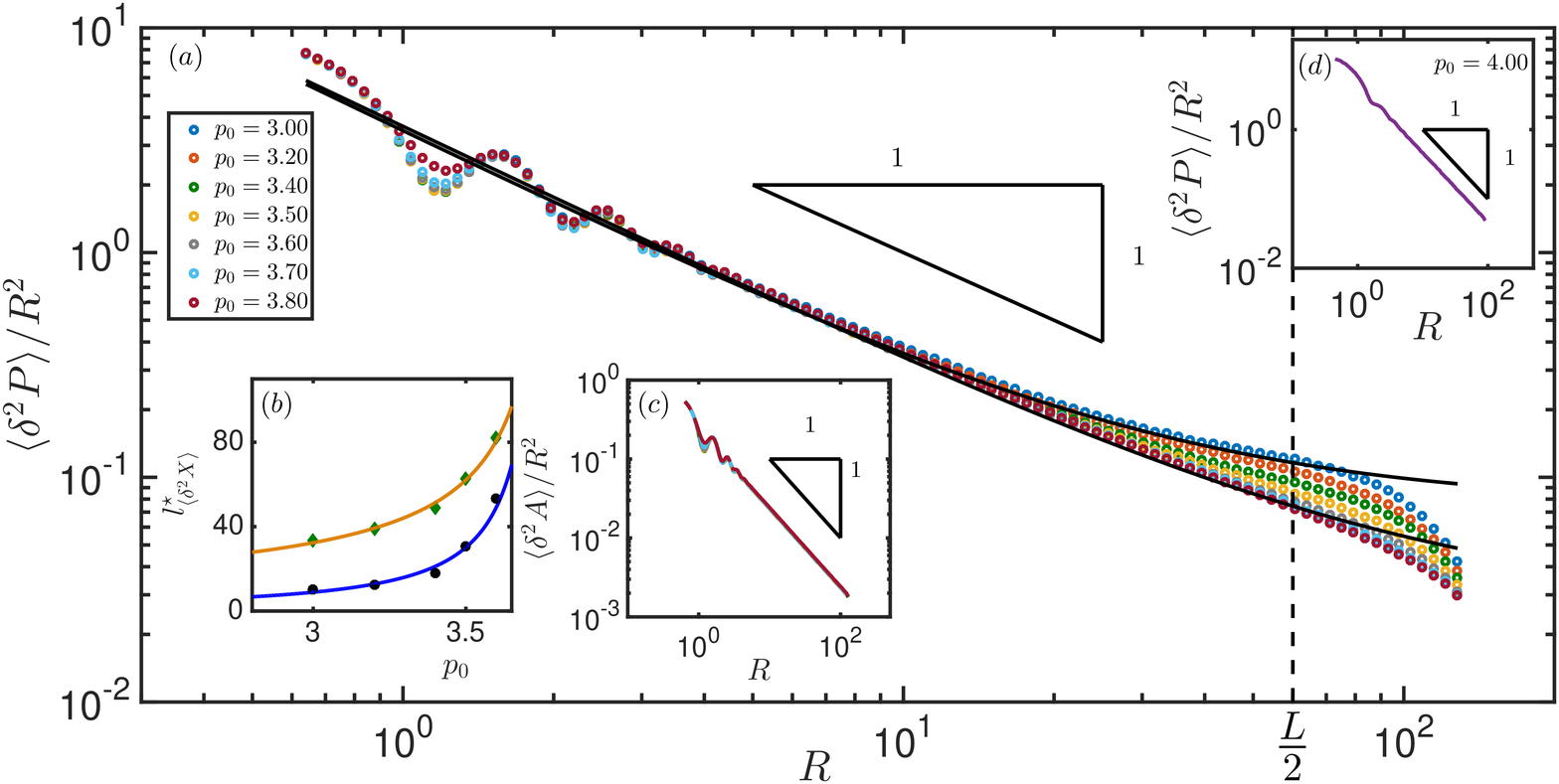}
\caption{ (a) Perimeter fluctuations $\langle \delta^2 P \rangle /R^2$,  as a function of $R$ at various values of $p_0$ in the rigid phase ($p_0 < p_0^{\star}$) for a system size of $N=16384$ particles. (b) $l^{\star}_{\delta^2 P}$  extracted from a fit of scaled perimeter fluctuations to the ansatz - $\langle \delta^2 P \rangle /R^2 \sim kx^{\alpha}+\beta$ (black solid lines), for $R < L/2$ where $l^{\star}_{\delta^2 P} \equiv (\beta/k)^{1/\alpha}$. The orange line represents a fit of the numerically robust regime (green diamonds) to  $l^{\star}_{\delta^2 P} \sim \vert \vert p_0-p_0^{\star}\vert \vert^{-\mu}$ where $\mu = 0.6824$, indicating an enhanced suppression of the perimeter fluctuations over its density counterpart.(c) $\langle \delta^2 A \rangle /R^2$ displays a linear scaling with $R$ for all values of $p_0$ in the rigid phase ($p_0 < p_0^{\star})$, a consequence of close packing constraints of the Voronoi tessellation. (d) Perimeter fluctuations are strongly suppressed in the fluid phase $(p_0=4.00)$} \label{fig:geometrical_fluctuations} 
\end{figure*}

The area fluctuations $\langle \delta^2 A \rangle$ are hyperuniform for all values of $p_0$ and $R$ due to strong correlations induced by the close packing constraint of the Voronoi tessellation (see Fig.\ref{fig:geometrical_fluctuations}(c)), such that the behavior in the thermodynamic limit is identical for any general 
distribution of points. However, the perimeter fluctuations exhibit a highly non-trivial behavior similar to that of the density - they are only suppressed up to a finite $p_0$ dependent length scale in the solid phase, and are hyperuniform in the liquid phase (see Fig. \ref{fig:geometrical_fluctuations}(a,d)).
By assuming a similar functional form for the scaling ansatz used previously in the analysis of the density fluctuations, we extract an analogous length scale $l_{\langle \delta^2 P\rangle}$ that is associated to the perimeter fluctuations. This lengthscale is again found to diverge at $p_0^{\star}$ but with a significantly less pronounced exponent $\nu_P \sim 0.68$ (Fig \ref{fig:geometrical_fluctuations}(b)). Thus, the suppression of perimeter extends over a much longer lengthscale than its density counterpart. 

We highlight that this is despite the fact that the underlying density fluctuations associated to the spatial distribution of points are themselves not suppressed for $ l_{\langle \delta^2 P\rangle} < R < l_{\langle \delta^2 N\rangle} $. This is unlike \cite{Torquato2016}, where fluctuations of random scalar fields defined on underlying point distributions that are hyperuniform by construction, are also found to be anomalously suppressed themselves. This implies that the system at hand must generate additional correlations in the perimeter field for the generalized hyperuniformity to surface. 

\begin{figure*}
	\includegraphics[width=\textwidth]{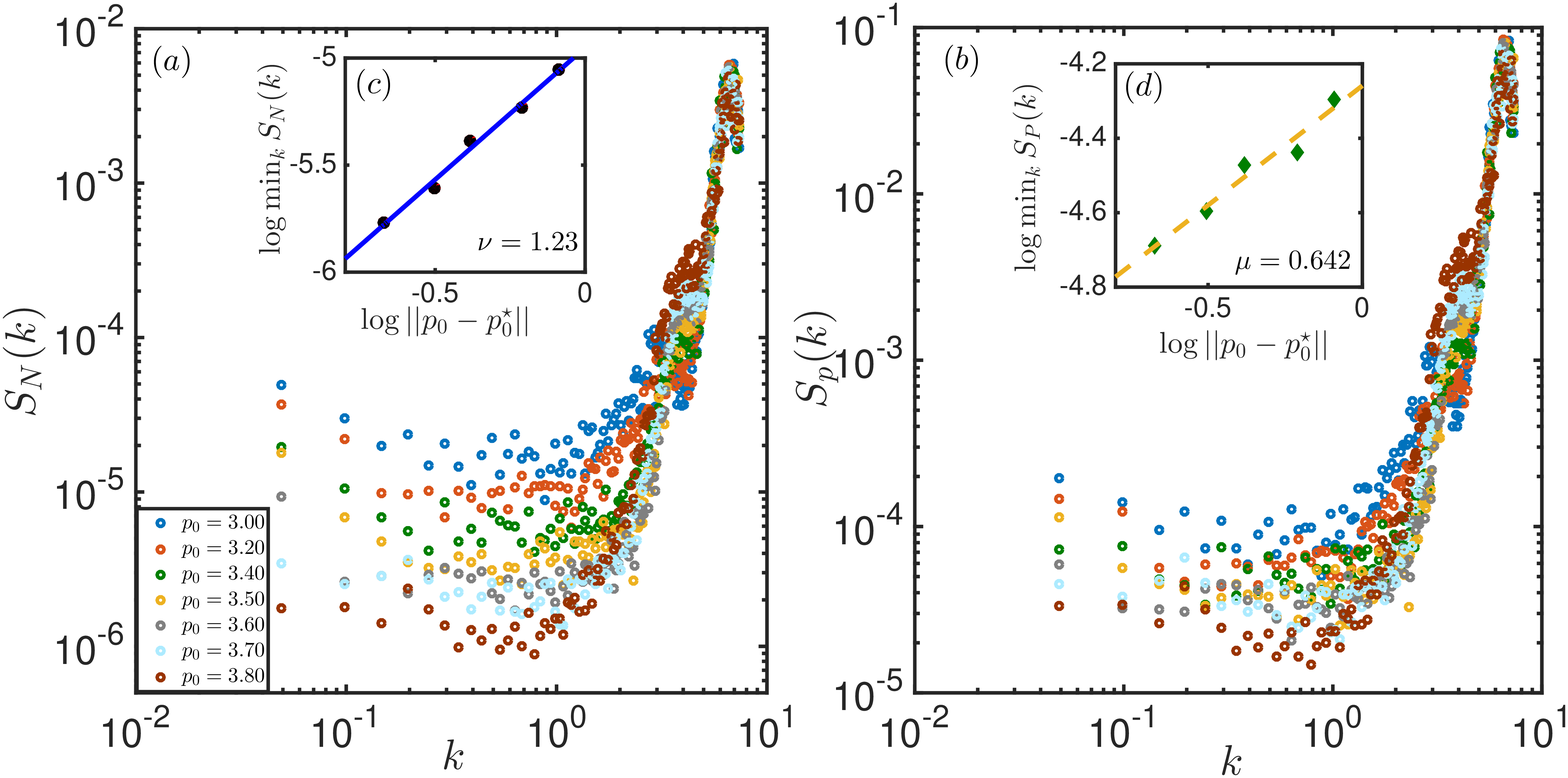}
	\caption{(a) $S_N(k)$ and (b) $S_P(k)$ as a function of $R$ at  various values of $p_0$ in the rigid phase ($p_0 < p_0^{\star}$) for a system size of $N=16384$ particles. (Inserts) Power law fits for the minimum value of $S_N(k)$ and $S_P(k)$ yields exponents $\nu=1.23$ and  $\mu=0.642$} 
	\label{fig:structure_factor}
\end{figure*}

 Similar estimates of the scaling exponents $\nu $ and $\mu$ are also recovered from the respective fits of the static structure factor $S_N(\vec{k})=\frac{1}{N}\left\langle \left\vert \sum_{i=1}^N e^{-i\vec{k}.\vec{r}_i} \right\vert \right\rangle$ and the perimeter generalized structure factor $S_P(\vec{k})=\frac{1}{N}\left\langle \left\vert \sum_{i=1}^N P_i e^{-i\vec{k}.\vec{r}_i} \right\vert \right\rangle$ (Fig. \ref{fig:structure_factor}), indicating that our results derived from real space are robust.

We draw attention and analogy of this observed behavior to the recent
observation that randomly jammed spheres suppresses the fluctuation in
$Z$ up to a length scale that diverges at the jamming transition~\cite{HexnerNagel2017},
while appearing not hyperuniform in their density~\cite{IkedaBerthier2015,IkedaParisi2017,GodfreyMoore2018,WuTeitel2015}, 
possibly due to the presence of rattlers. 
In isostatic sphere packings, the average number of contacts per particle is fixed, $Z = 2d$ in $d$ spatial dimensions. Likewise, in the present model, at the transition and in the fluid phase the perimeter of particles is fixed such that $P_i = p_0$. Accordingly, what emerges seems to be an enhanced suppression of fluctuations in quantities 
that directly correspond to the mechanical stability of the system.

Considered collectively, these various results in the different models indicate 
that complex systems are capable of self organizing in preference of suppressing 
fluctuations in quantities that goes well beyond what is indicated by its 
underlying density fluctuations. This also implies the presence of additional 
hidden long range correlations in quantities that may hint at a more fundamental 
and physical origin to the observed density fluctuation suppression at shorter 
length scales. From this perspective, the various density hyperuniformity 
observed at the Jamming transition in certain models, even if robust, could be 
one of coincidence and not causal. Lastly, we also highlight that frustration in 
the system is also crucial to the resultant fluctuation profiles observed. In 
the scenario where there is a lack thereof frustration, for instance in the 
fluid state of this model, the system is able to simultaneously suppress all 
fluctuations, leading to a strong hyperuniform behavior in all quantities. 

In this work, we investigated notions of hyperuniformity in a 
model known to exhibit a rigidity transition. Unlike traditional models of 
Jamming, the inherent manybody interactions of the Voronoi model by 
construction, circumnavigates the appearance of pairwise quantities of the 
underlying degrees of freedom in the energy functional, such that subtle 
connections to hyperuniformity and generalized fluctuations are revealed. In the solid phase 
we find an anomalous suppression of density fluctuations up to a finite length scale,
contrary to earlier findings. This lengthscale grows as we approach the rigidity transition 
and diverges in the fluid phase, where the system is thus hyperuniform in its density.
Most surprisingly, we find that fluctuations of the perimeter are suppressed beyond 
what is indicated by its density fluctuations, alluding to the presence of 
additional correlations and self-organization that is driven by an underlying 
physical principle that still remains largely hidden. 
We speculate that jammed systems suppress the fluctuations of quantities that
are directly correlated with the condition of mechanical stability.

We acknowledge support from the Singapore Ministry of Education through the
Academic Research Fund MOE2017-T2-1-066 (S), and are grateful to the National Supercomputing Centre (NSCC) for providing computational resources.
 
\newpage
\beginsupplement
\section{Supplementary material}
\begin{figure}[!htb]
	\includegraphics[width=\columnwidth]{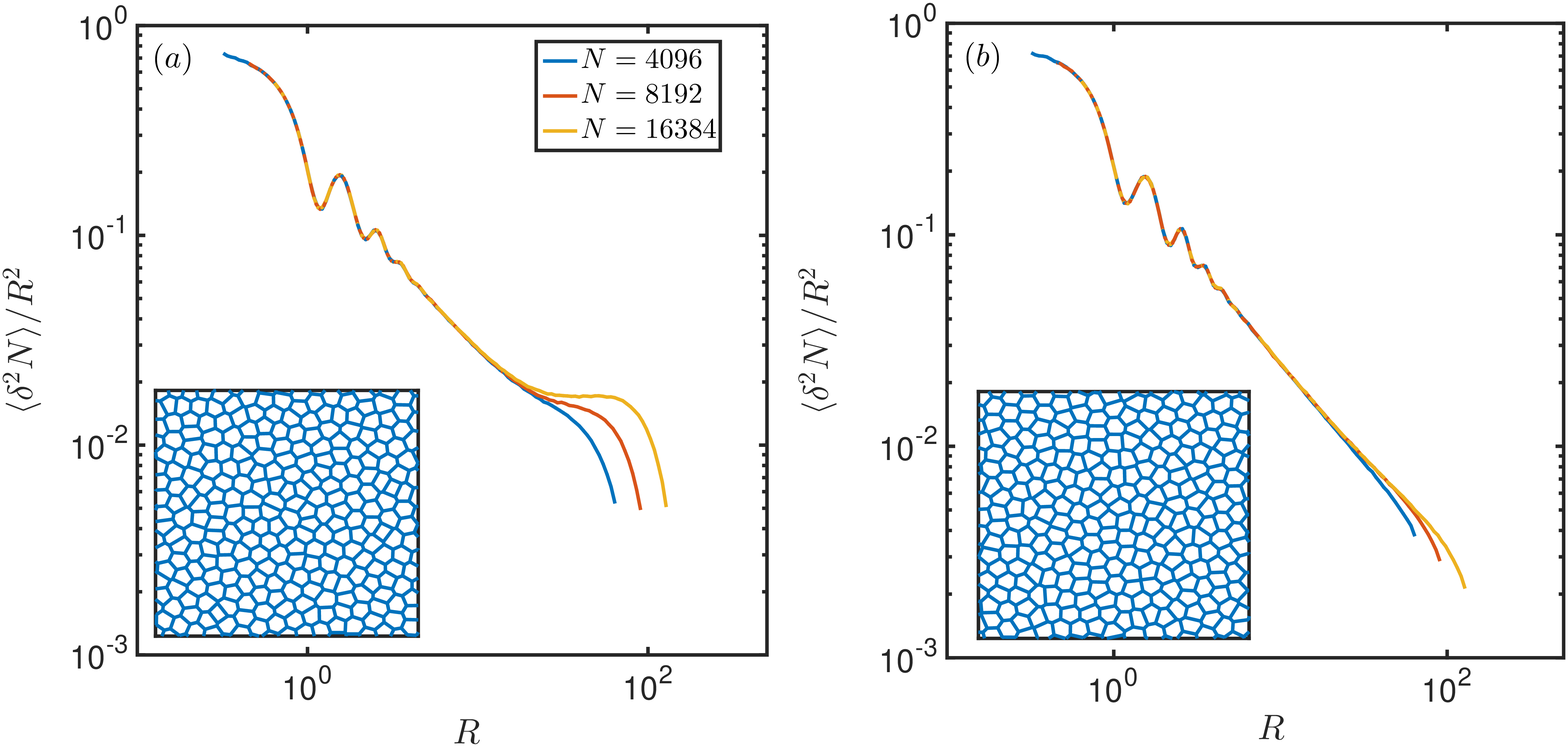}
	\caption{ $\langle \delta^2 N \rangle /R^2$ 
for various system sizes $(N=4096, N= 8192,  N=16384)$ with $\bar{A}=1$ at (a) 
$p_0=3.20$ and (b) $p_0=3.80$. The crossover behavior in $\langle \delta^2 N 
\rangle /R^2$ is masked for smaller systems in which the physical length 
$L=\sqrt{N}$ is significantly smaller than the density lengthscale 
$l^{\star}_{\delta^2 N}$. (Inserts) Snapshots of sub-regions about the size of 
$l^{\star}_{p_0=3.20}$ for the respective values of $p_0$ for $N=4096$. }
	\label{fig:finite_size}
\end{figure}
\end{document}